\documentstyle[epsf]{mn}

\newcommand{\subscr}[1]{_{\rm #1}}

\newcommand{\Mdot}{\dot{M}}
\newcommand{\Msol}{\rm M_{\odot}}

\title{On the migration of a system of protoplanets}
\author[W. Kley]
       {W. Kley\\
       Theoretisch-Physikalisches Institut,
       Universit\"at Jena, Max-Wien-Platz~1, D-07743 Jena, Germany\\
       }

\begin{document}
\maketitle
\epsfverbosetrue
\begin{abstract}
The evolution of a system consisting of a protoplanetary disc with
two embedded Jupiter sized planets is studied numerically.
The disc is assumed to be flat
and non-self gravitating, which is modeled by the planar (two-dimensional)
Navier-Stokes equations.  
The mutual gravitational interaction of the planets and the star,
and the gravitational torques of the disc acting
on the planets and the central star are included.
The planets have an initial mass of one Jupiter mass $M_{Jup}$
each and the radial
distances from the star are one and two semi-major axis of Jupiter,
respectively.

During the evolution both planets increase their mass due to
accretion of gas from the disc; after about 2500 orbital periods of the
inner planet it has reached a mass of 2.3 and the outer planet of
3.2~$M_{Jup}$.
The net gravitational torques exerted by the disc on the planets
result in an inward migration of the outer planet on time-scales comparable
to the viscous evolution time of the disc, while the semi-major axis
of the inner planet remains constant.
When the distance of close
approach eventually becomes smaller than their mutual Hill radius the
eccentricities increase strongly and the system may turn unstable.

The implications for the origin of the solar system and the extrasolar
planets are discussed.
\end{abstract}
\section{Introduction}
It is generally assumed that planetary systems form in a differentially
rotating gaseous disc.
In the late stages of their formation the protoplanets
are still embedded in the protostellar disc and their orbital
evolution is coupled to that of the disc.
Gravitational interaction between the planets and the gaseous
disc has basically two effects:
\\a)
The torques by the planets acting on the disc tend to push
away material from the orbital radius of the planet, and for sufficiently
massive planets a {\it gap} is formed in the disc (Papaloizou \& Lin 1984;
Lin \& Papaloizou 1993).
The dynamical process of gap formation has been studied through time
dependent hydrodynamical simulations for planets on circular orbits
by Bryden et al. (1999) and Kley (1999, henceforth paper I).
The results indicate that even after the formation of a gap, the planet
may still accrete material from the disc and reach about 10 Jupiter masses
($M_{Jup}$). For very low disc viscosity and larger planetary masses 
the mass accumulation finally terminates (Bryden et al. 1999).
\\b) 
Additionally, the gravitational force exerted by the disc alters 
the orbital parameter (semi-major axis, eccentricity) of the
the planet. Here, these forces typically induce some {\it inward migration}
of the planet (Goldreich \& Tremaine 1980) which is coupled
to the viscous evolution of the disc (Lin \& Papaloizou 1986). 
Hence, the present location of the observed planets (solar and extrasolar)
may not be identical with the position at which they formed.

In particular, the migration scenario applies to some of the extra-solar
planets (for a summary of their properties see 
Marcy, Cochran \& Mayor 1999), the 51~Peg-type planets.
They all have masses of the order $M_{Jup}$,
and orbit their central stars very
closely, having orbital periods of only a few days.
As massive planets, according to standard theory, 
have formed at a few AU distance from their stars,
these planets must have migrated to their present position.
The inward migration was eventually halted by tidal interaction
with the star or through interaction with the stellar magnetosphere
(Lin, Bodenheimer \& Richardson 1996).
The only extrasolar planetary system known
so far ($\upsilon$ And) consists of one planet at 0.059 AU on a 
nearly circular orbit
and two planets at .83 and 2.5 AU having larger eccentricities (.18 and .41)
(Butler et al. 1999).

In case of the solar system the question,
what prevented any further inward migration of Jupiter, arises.
As the net tidal torque on the planet is a delicate balance between
the torque of the material located outside of the planet and the material
inside (eg. Ward 1997),
any perturbation in the density distribution may change this balance.
In this letter we consider the effect that an additional planet
in the disc has on the migration rate.

We present
the results of numerical calculations of a thin, non-self gravitating,
viscous disc with two embedded protoplanets.
Initially the planets with one $M_{Jup}$ each are on circular orbits at  
$a= 1 a_{Jup}$ and 2 $a_{Jup}$, respectively.
In contrast to the existing time-dependent models (Bryden et al. 1999;
paper I) we take into account the back-reaction of the gravitational force
of the disc on the orbital elements of the planets and star.
The models are run for about 3000 orbital periods of the inner planet
corresponding to 32,000 years.
In Section 2 a description of the model is given, the results are presented
in Section 3 and our conclusions are given in Section 4. 

\section{The Model}
\noindent 
We consider a non-self-gravitating, thin accretion disc model for the
protoplanetary disc located in the $z=0$ plane and
rotating around the $z$-axis. Its evolution is
described by the two dimensional
($r-\varphi$) Navier-Stokes equations, which are given in detail
in Kley (1999, paper I). 
The motion of the disc takes place in the gravitational field of
the central star with mass $M_s$ and the two 
embedded protoplanets with masses $m_1$ and $m_2$. 
In contrast to paper I we use here 
a non-rotating frame as both planets have to be moved through the grid.
The gravitational potential is then given by
\begin{equation}  \label{Phi}
    \Phi = - \frac{G M_s}{| {\bf r} - {\bf r}_s |}
       -  \frac{G m_1}{ \left[ ( {\bf r} - {\bf r}_1 )^2 
             + s_{1}^2 \right]^{1/2} }
       -  \frac{G m_2}{ \left[ ( {\bf r} - {\bf r}_2 )^2 
             + s_{2}^2 \right]^{1/2} }
\end{equation}
where $G$ is the gravitational constant and ${\bf r}_s$,
${\bf r}_1$, and ${\bf r}_2$ are the radius vectors to the star and two
planets, respectively.
The quantities $s_1$ and $s_2$ are smoothing lengths
which are 1/5 of the corresponding sizes of the Roche-lobes.
This smoothening of the potential allows the motion of the planets through
the computational grid.

The motion of the star and the planets is determined firstly by their
mutual gravitational interaction and secondly by the gravitational
forces exerted on them by the disc.
The acceleration of the star ${\bf a}_s$ is given for example by
\begin{eqnarray}  \label{a_s}
    {\bf a}_s = 
       -  G m_1 \frac{ {\bf r}_s - {\bf r}_1}{ | {\bf r}_s - {\bf r}_1 |^3}
     &  - &  G m_2 \frac{ {\bf r}_s - {\bf r}_2}{ | {\bf r}_s - {\bf r}_2 |^3}
         \nonumber \\
     &  - & G \int_{Disc} \Sigma \, \frac{ {\bf r} - {\bf r}_s}
              { | {\bf r} - {\bf r}_s |^3} \,  dA
\end{eqnarray}
where the integration is over the whole disc surface, and $\Sigma$ denotes
the surface density of the disc.
The expressions for the two planets follow similarly.
We work here in an accelerated coordinate frame where the origin
is located in the centre of the (moving) star.
Thus, in addition to the gravitational potential (\ref{Phi}) the disc
and planets feel the additional acceleration $-{\bf a}_s$. 

The mass accreted from the disc by the planets (see below)
has some net angular momentum which in principle changes also the
orbital parameter of the planets. However, this contribution is typically
about an order of magnitude smaller than the tidal torque
(Lin et al. 1999) and is neglected here.

As the details of the origin and magnitude of the viscosity in discs
is still uncertain we assume a Reynolds-stress formulation (paper I)
with a {\it constant} kinematic viscosity.
The temperature distribution of the disc is fixed throughout the computation
and is given by the assumption of a constant ratio of the vertical
thickness $H$ and the radius $r$. 
Hence, the fixed temperature profile is given by $T(r) \propto r^{-1}$.
We assume $H/r = 0.05$, which is equivalent to a fixed Mach number of 20.

For numerical convenience we introduce dimensionless units, in which the
unit of length, $R_0$, is given by the initial distance of the first planet
to the star $R_0 = r_1 (t=0) = 1 a_{Jup}$.
The unit of time is obtained from the (initial) orbital angular frequency 
$\Omega_1$ of the first planet
i.e. the orbital period of the planet 1 is given by 
\begin{equation}
   P_1 = 2 \, \pi t_0  \label{P1}.
\end{equation}
The evolutionary time of the results of the calculations as given below
will usually be stated in units of $P_1$. The unit of velocity is then
given by $v_0 = R_0 / t_0$.
The unit of the kinematic viscosity coefficient is given by
$\nu_0 = R_0 v_0$. Here we take a typical dimensionless value of $10^{-5}$
corresponding to an effective $\alpha$ of $4 \times 10^{-3}$.

\subsection{The numerical method in brief}
The normalized equations of motion are
solved using an Eulerian finite difference scheme,
where the computational domain $[r_{min}, r_{max}] \times
[\varphi_{min}, \varphi_{max}]$
is subdivided into $N_r \times N_\varphi$
grid cells.
For the typical runs we use $N_r = 128, N_\varphi = 128$,
where the azimuthal spacing is equidistant, and the radial
points have a closer spacing near the inner radius.
The numerical method is based on a spatially second order accurate
upwind scheme (monotonic transport), which uses a formally first
order time-stepping procedure. The methodology of the finite difference
method for disc calculations
is outlined in Kley (1989) and paper I.

The N-body module of the programme uses a forth order Runge-Kutta
method for the integration of the equations of motion. It
has been tested for long term integrations using the onset of instability
in the 3-body problem consisting
of two closely spaced planets orbiting a star as described by
Gladman (1993). 
For the initial parameter used here, the error in the total
energy after $1.2 \times 10^5$ orbits 
(integration over $10^6 yrs$) is less $2 \times 10^{-9}$. 

\subsection{Boundary and initial conditions} \label{bounds}
To cover the range of influence of the planet on the disc
fully, we typically choose for the radial extent
(in dimensionless units, where planet 1 is located initially at $r=1$)
$r_{min} = 0.25, r_{max} = 4.0$. 
The azimuthal range covers a complete ring
$\varphi_{min} = 0.0, \varphi_{max} = 2 \pi$
using periodic boundary conditions.
To test the accuracy of the migration, a comparison calculation with
$r_{max}=8.0$ and higher resolution
$N_r = 256, N_\varphi = 256$ was also performed.

The outer radial boundary is closed to any mass flow $v(r\subscr{max})=0$,
while at the inner boundary mass outflow is allowed, emulating accretion
onto the central star.
At the inner and outer boundary the angular velocity is set to the value
of the unperturbed Keplerian disc.
Initially, the matter in the domain is distributed axially symmetric with
a radial surface density profile $\Sigma \propto r^{-1/2}$.

Two planets, each with an initial mass of $1 M_{Jup}$,
are located at $r_1=1.0, \varphi_1=\pi$ and $r_2=2.0, \varphi_2=0$.
Thus, they are not only spaced in radius but are positioned (in $\varphi$)
in opposition to each other to minimize the initial disturbance.
The radial velocity $v$ is set to zero, and the angular velocity is
set to the Keplerian value of the unperturbed disc.

Around the planets we then introduce an initial density reduction whose
approximate extension is obtained from their masses and
the magnitude of the viscosity.
This initial lowering of the density is assumed to be axisymmetric; the
radial profile $\Sigma(r)$ of the initial distribution
is displayed in Fig.~1 (solid line).
The total mass in the disc depends on the physical extent of the computational
domain. Here we assume a total disc mass within $r_{min}=0.25$ and
$r_{max}=4.00$ of 0.01 $\Msol$.
The starting model is then evolved in time and the accretion rates onto
the planets is monitored, where a given fraction of the mass
inside the Roche-lobe of the planet is assumed to accrete onto the planet
at each time step and is taken out of the computation and added to
masses of the planets.
The Courant condition yields a time step of $6.8 10^{-4} P_1$.

\section{Results}
%
\begin{figure}
\epsfxsize=8.5cm
\epsfbox{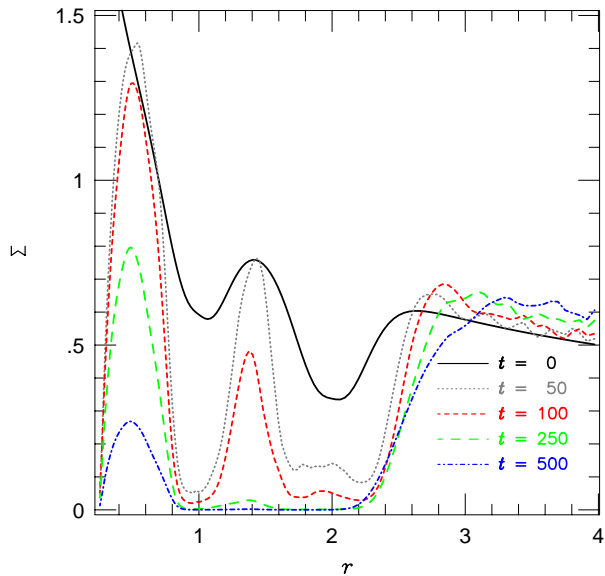}
  \caption{The azimuthally averaged surface density for different times
      (stated in units of $P_1$).
     The solid line refers to the initial density distribution.
     The density inside planet 1 is reduced because of mass leaving
     through $r_{min}$. 
     }
\end{figure}
%
\begin{figure}
\epsfxsize=8.5cm
\epsfbox{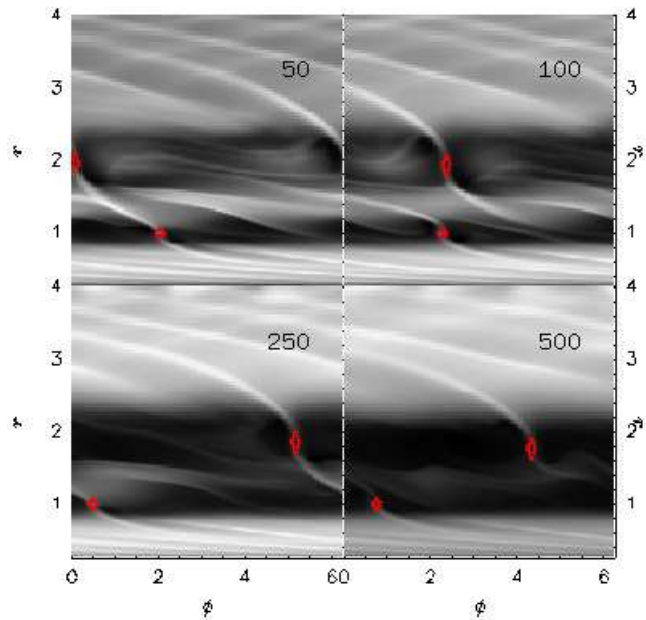}
  \caption{Grey-scale plot of the surface density of the at four
      different times in an $\varphi - r$ diagram. The small ellipses 
     indicate the positions of the planets and sizes of their Roche lobes.}
\end{figure}

Starting from the initial configuration (Fig.~1) the planets exert torques
on the adjacent disc material which tend to push mass away from the
location of the planets. At the same time the planets continuously
accrete mass from its surroundings, and the mass
contained initially between the two planets is
quickly accreted by the planets and added to their mass.
Finally one large gap remains in the region between $r=1$ and $r=2$
(Fig.~1).

Similarly to the one planet calculations as described in Bryden et al.
(1999) and in paper I, each of the two planets
creates a spiral wave pattern (trailing shocks) in the density of the disc. 
In case of one disturber on a circular orbit
the pattern is stationary in the frame co-rotating with the planet.
The presence of a second planets makes the spirals non-stationary as is
seen in the
snapshots after 50, 100, 250 and 500 orbits of the inner planet that are
displayed in figure 2. Near the outer boundary at $r=4$ the reflection of the
spiral waves are visible. Using the higher resolution model 
with $r_{max}=8.0$ (section /ref{bounds})
we tested whether the numerical resolution or the wave reflection
at $r_{max}$ has any influence
on the calculation of the net torques acting on the planet
and the accretion rates onto the planet. 
Due to limiting computational resources the higher resolution model
was run only for a few hundred 
orbital periods and the largest difference ($2.5\%$)
occurred in the mass $m_3$ of the outer planet. The difference in radial
position (migration) is less than $1\%$. We may conclude that our
resolution was chosen sufficiently fine and that the reflections at the
outer boundary $r_{max}=4$ do not change our conclusions significantly.

In previous calculations (paper I) the equilibrium
mass accretion rate from the outer part of the disc 
onto a one Jupiter mass planet for the same viscosity
($\nu = 10^{-5}$) and distance from the star
was found to be $4.35 \times 10^{-5} M_{Jup}/yr$
for a fully developed gap.
Here the accretion rate onto the planets is much higher in the beginning 
($\approx 5 \times 10^{-4} M_{Jup}/yr$) as the initial
gap was not as cleared. Thus, during this gap clearing process,
the masses of the
individual planets grows rapidly at the onset of calculations (Fig.~3).
At $t\approx 250$ the mass within the the gap has been exhausted (see also
Fig.~1) and the accretion rates $\Mdot$ on the planets lower dramatically.
At later times after several hundred orbits ($P_1$) they settle
to nearly constant values of about $10^{-5} M_{Jup}/yr$ for the outer planet,
and $2.2 \times 10^{-6} M_{Jup}/yr$ for the inner
planet (from Fig.~3). Since the mass inside of planet 1 has left the
computational domain and the initial mass between the two
planets has been consumed by the two planets, this mass accretion rate onto
planet 1 for later times represents the mass flow of material coming from
radii larger than $r_2$ (beyond the outer planet).
It is the material which has been flowing {\it accross} the gap of the outer
planet.
Previously, this mass flow across a gap has been calculated to be about
one seventh of the mass accretion rate onto a planet (paper I) and the
present results are entirely consistent with that estimate.

\begin{figure}
\epsfxsize=8.5cm
\epsfbox{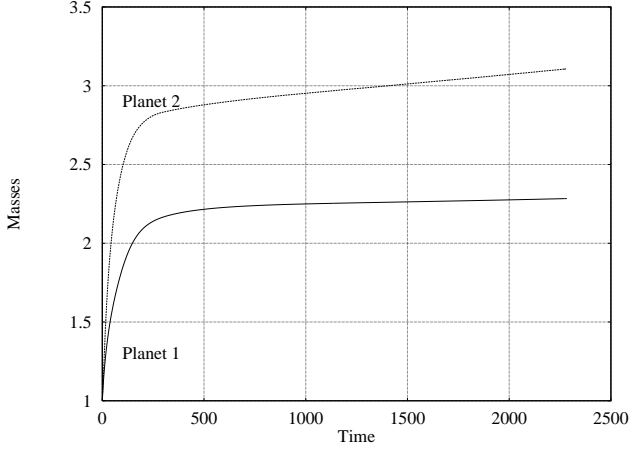}
  \caption{The increase of the masses of the planets in units of $M_{Jup}$.
     }
\end{figure}
The gravitational torques exerted by the disc lead
to an additional acceleration of the planets resulting in an expression
similar to
the acceleration of the star (Eq.~\ref{a_s}). For one individual planet
this force typically results in an inward migration of the planet
on timescales of the order of the viscous time of the disc 
(Lin \& Papaloizou 1986). 
Here this inward migration is seen clearly for the outer planet
in Fig.~4 where the time evolution of
the semi-major axis of the two planets is plotted.  

The inner planet on the other hand initially, for $t<200$, moves
slightly inwards but then the semi-major axis increases and,
showing no clear sign of migration anymore, settles to
a constant mean value of $1.02$.
However, the decrease of the semi-major axis of the outer planet reduces
the orbital distance between the two planets. From three body simulations
(a star with two planets) and analytical considerations
(Gladman 1993; Chambers, Wetherill \&  Boss 1996) it is known that
when the orbital distance of two planets lies below the critical
value of $\Delta_{cr} = 2 \sqrt{3} R_{H}$, where
\begin{equation}
     R_H  = \left( \frac{m_1 + m_2}{3 M_*} \right)^{1/3} \frac{a_1 + a_2}{2}
\end{equation}
is the mutual Hill radius of the planet,
the orbits of the planets are not stable anymore.
In the calculations this effect is seen in the strong increase of the
eccentricity of the inner planet.
At $t=2500$ its eccentricity has grown to about 
$e_1 = 0.1$, while the eccentricity of the outer planet remains approximately
constant at a level of $e_2 = 0.03$.

We should remark here that in the pure 3-body problem
without any disc and the same initial conditions ($r_1=1.0, m_1=1.0;
r_2=2.0, m_2=1.0$) for the three objects,
the semi-major axis of the planets stay constant as this system is
definitely Hill stable (Gladman 1993).
However, if one takes as initial conditions for the pure 3-body system
the parameters for the planets as obtained from the disc evolution at
$t=2500$ ($r_1=1.0, m_1=2.3; r_2=1.5, m_2=3.2$) then the evolution
becomes chaotic on timescales of hundreds of orbits and the
eccentricity grows up to $e=0.6$ for both planets within
$4000$ orbits.
\begin{figure}
\epsfxsize=8.5cm
\epsfbox{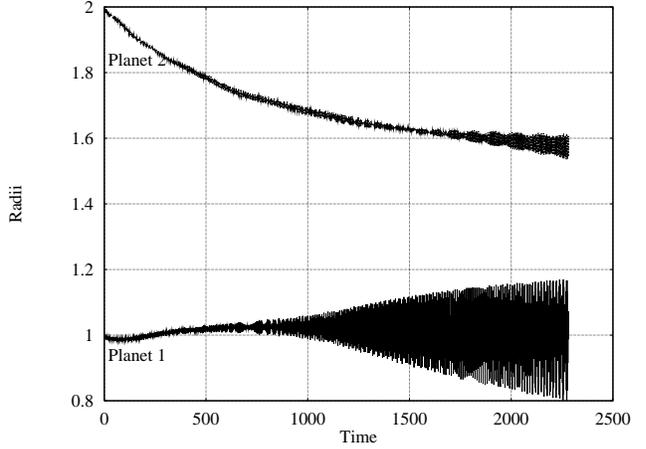}
  \caption{The evolution of the semi-major axis of the planets.
     }
\end{figure}
\section{Conclusions}
We have presented calculations of the long term
evolution of two embedded planets
in a protoplanetary disc that covers several thousand orbital periods.
The planets were initially located at one and two $a_{Jup}$ from the central
star with initial masses of $1 M_{Jup}$ each.
The gravitational interaction with the gaseous disc having
a total mass of $0.01 \Msol$ leads to an inward migration of the outer
planet while the semi-major axis of the inner planet remains approximately
constant and even slightly increased.
At the same time, the ongoing accretion onto the planets increases their
masses continuously until at the end of the simulation (at $t=2500$) the
outer planet has reached a mass of about $3.2 M_{Jup}$ and the inner planet
of about $2.3 M_{Jup}$.

This increase in mass and the decreasing distance between them renders
the orbits eventually unstable resulting in a strong increase of the
eccentricities on timescales of a few hundred orbits.

From the computations we may draw three major conclusions:
\\1) The inward migration of planets immersed in an accretion disc
may be halted by the presence of additional protoplanets located
for example at larger radii. They
disturb the density distribution significantly which in turn
reduces the net gravitational torque acting on the inner
planet.
Thus, the migration of the inner planet is halted, and its
semi-major axis remains nearly constant.
\\2) When disc depletion occurs sufficiently rapid to prevent
a large inward migration of the outer planet(s),
a planetary system with massive planets
at a distance of several $au$ remains. This scenario may explain why for
example in the solar system the outer planets (in particular Jupiter)
have not migrated any closer towards the sun.
\\3) If the initial mass contained in the disc is sufficiently
large then the inward
migration of the outer planet(s) will continue until some of them
reach very close spatial separations. This will lead to unstable orbits
resulting in a strong increase of the eccentricities. Orbits may
cross and planets may then be driven either to highly eccentric orbits
or may leave
the system all together (see eg. Weidenschilling \& Marzari 1996).
This may then explain the high eccentricities in some of the
observed extrasolar planets in particular the
planetary system of $\upsilon$~Andromedae.

As planetary systems containing more than two planets display different
stability properties (Chambers et al. 1996) it will be interesting to
study the evolution of multiple embryos in the protoplanetary nebula.
\section*{Acknowledgments}
I would like to thank Dr F. Meyer for lively discussion on this topic.
Computational resources of the Max-Planck Institute for Astronomy
in Heidelberg were available for this project and are gratefully
acknowledged.  This work was supported by the Max-Planck-Gesellschaft,
Grant No. 02160-361-TG74.

\end{document}